\documentclass[twocolumn,amssymb,aps,pra,showpacs]{revtex4-1}
\usepackage{amsmath}
\usepackage{amssymb}
\usepackage{ifsym}
\usepackage[colorlinks,linkcolor=blue,anchorcolor=blue,citecolor=blue,urlcolor=blue,dvipdfm]{hyperref}
\usepackage{graphicx}
\begin{document}
\title{Rydberg states generation of Hydrogen atoms with intense laser pulses:
the roles of Coulomb force and initial lateral momentum}

\author{Bin Zhang}
\author{Wenbo Chen}
\author{Zengxiu Zhao}
\email{zhao.zengxiu@gmail.com}

\affiliation{Department of Physics, College of Science, National University of Defense Technology, Changsha, Hunan 410073, China}

\date{\today}
%\maketitle

%\thispagestyle{empty}

\begin{abstract}
We investigate the Rydberg states generation of Hydrogen atoms with intense laser pulses, by solving
the time-dependent Schr\"odinger equation and by means of classical trajectory monte-carlo simulations.
Both linearly polarized multi-cycle pulses and pairs of optical half cycle pulses are used.
Comparisons between these methods show that both the Coulomb force and initial lateral momentum, which have effects on the $n$-distribution and $l$-distribution of the population of excited states, are important in the generation of Rydberg states.
\end{abstract}


\maketitle

%\pagenumbering{roma}
%%%%%%%%%%%%%%%%%%%%%%%%%%%%%%%%%%%%%%%%%%%%%%%%%%%%%%%%%%%%%%%%%%%%%%%%%%%%%%

\vspace{8mm}

%\newpage \pagenumbering{arabic}
\section{INTRODUCTION}

Tunnel ionization is a fundamental atomic and molecular process in strong laser fields~\cite{Ammosov86}.
The tunneled electron is accelerated in the fields and may return to the vicinity of the ion, resulting in many highly non-linear strong-field phenomena, such as:
high-harmonic generation, above-threshold ionization, and non-sequential double ionization~\cite{krausz09}.
Recent experiment showed that in the tunneling regime of strong-field ionization of Helium atoms, a substantial fraction of neutral atoms survived the laser pulse in excited states, which was explained with the strong-field tunneling-plus-rescattering model and named frustrated tunneling ionization(FTI)~\cite{Nubbemeyer08}.
The FTI is a completion to the tunneling-rescattering scenario.

The three-step model~\cite{Corkum93}, which neglects the Coulomb force after ionization, works surprisingly well in explaining many phenomena, most notably high-harmonic generation.
For the generation of Rydberg states, a semiclassical model that neglected Coulomb force during the laser pulse was used to explain
experimental observations in the setting of elliptically polarized light~\cite{Landsman13}.
However, more recently, the role of the Coulomb force in laser ionization has drawn considerable attention in many studies~\cite{Liu12,wu12,zhang13,Muller99},
where the inclusion of the Coulomb force is proven to be essential.
While some progress has been made in analyzing the generation of Rydberg states
in both elliptically~\cite{Landsman13} and linearly polarized light~\cite{Shilovskia09},
the underlying physics, especially the role of Coulomb force,
still remain to be explained.
We will also investigate the role of the initial lateral momentum distribution, as accompanied in the tunneling process~\cite{delone91}.

Previous investigations focused on the generation of Rydberg atoms with multi-cycle laser pulses~\cite{Landsman13,Shilovskia09,Preclikova12,Vrijen97}.
In this paper, we first present some results utilizing the multi-cycle laser pulses.
To filter out the complexity and gain better insight into the underlying mechanism, we have also utilized a pair of optical half cycle pulses (HCP).
The HCP fields in the frequency of Tera-hertz (THz) regime have been widely used in the investigation
and controlling the Rydberg states in the past~\cite{Wetzels02,Wesdorp01,Wetzelsa01,Jones96,Noordam92,Bensky97,Raman96,Hu04}.
If the pulse duration of a HCP is very short as compared to the
orbital time of a Rydberg electron, the impact
of this HCP is generally described as a momentum kick~\cite{Wetzels02}.
No works have been done using the HCPs in the frequency of optical regime.
Such optical HCPs have several advantages against THz HCPs:
(1)Due to the weak intensity, the THz HCP can only be used to generate a THz Rydberg wavepacket
(a superposition of the initial state and its neighboring states)~\cite{Wetzelsa01} from an initial optically excited atom;
(2)The pulse duration of THz HCP is of (or near) the same order as the orbital time of a Rydberg electron, thus mixing the interaction with more complex wavepacket dynamics and making the "kick" description less valid;
(3)For the ground and low-lying excited states, the interaction with THz HCP is ab-initially difficult to treat since the electron moves in an atto-second time scale~\cite{krausz09}.
The theoretical investigation of Rydberg states generation with optical HCPs are thus meaningful and may provides
insides into the physics behind.

The organization of this paper is as follows:
In Sec. II we briefly describe the methods of solving the 3D time-dependent Schr\"odinger equation (TDSE) and
classical trajectory monte-carlo (MC) simulation.
Sec. III gives the calculation details.
In Sec. IV, we present our results and discussions in detail.
We conclude in Sec. V.

\section{THEORY}

\subsection{TDSE}

The TDSE for atomic Hydrogen in the presence of external laser fields [${\bf F}(t)$] can be
written as [atomic units (a.u.) are used unless otherwise stated],
\begin{equation}
\label{e1} i\frac{\partial \Psi({\bf r},t)}{\partial t}=\Big [
H_0+V({\bf r},t)\Big ]\Psi({\bf r},t),
\end{equation}
where the field-free Hamiltonian $H_0=-\nabla^2/2-1/r$ and laser-atom interaction $V({\bf r},t)={\bf F}(t)\cdot {\bf r}$.
The orbital $\Psi({\bf r},t)$ is expanded in the spherical harmonics,
\begin{equation}
\label{ewfexpand}
\Psi({\bf r},t)=\sum_{l=0}^{l_{max}}\sum_{m=-l}^{m=l}\frac{\phi_{lm}(r,t)}{r}Y_{lm}(\theta,\varphi).
\end{equation}
For linearly polarized laser fields, the expansion only consists the $m=0$ partial waves.

For the $r$ coordinate, we use the DVR basis functions.
To this purpose, the variable $r$ is first truncated from the semi-infinite $(0,\infty)$ domain into finite domain
$(0,r_{max}]$ (with $r_{max}$ sufficiently large).
The $r$ coordinate is then discretized using the generalized Gauss-quadrature points $r^i$ and weights $w^r_i$:
\begin{equation}
\left\{  \begin{aligned}
r_i &=L\frac{1+x_i}{1-x_i+\alpha},\\
w^r_i &=r^{'}_i w_i,
\end{aligned} \right.
\end{equation}
where the points and weights [\{$x_i$, $w_i$\}, i=1,...,$n_r$] are associated with the standard ($n_r$) points Gauss-Radau quadrature ($x_{n_r}=1$)~\cite{nc3rd}. $L$ and $\alpha=2L/r_{max}$ are the mapping parameters. This mapping function $r(x)$ allows for denser grids near the origin, leading to more accurate eigenvalues and
eigenfunctions~\cite{tong97}. The Coulomb singularity at the origin is avoided
since $r_{min}=r(x_1)>0$.

The radial partial wave $\phi_{lm}(r,t)$ is expanded in a product basis of
functions,
\begin{equation}
\phi_{lm}(r,t)=\sum_{i} c_{i}^{lm}(t) g_i(r),
\end{equation}
where the DVR functions read,
\begin{equation}
\label{eqdn2}
g_i(r)=\frac{1}{\sqrt{w^{r}_i}}\prod_{j\ne i}^{n_{r}}\frac{r-r_j}{r_i-r_j}.
\end{equation}
Note that the factor $1/\sqrt{w^{r}_i}$ is built in to remove the integration overlaps, which results
in the orthonormal condition, $\int_{r_{min}}^{r_{max}}g_i(r)g_j(r)dr=\delta_{ij}$~\cite{zhang12}.

The ground and excited states of Hydrogen atom are calculated by
dialogizing the ground state hamiltonian. Eq.~(\ref{e1}) is propagated
in time by the second order split-operator technique~\cite{zhang12}.
An absorbing layer between $r_b$ and $r_{max}$ is used to smoothly brings down the wave function
and prevents the un-physical reflection from the boundary.
After the time propagation, we get the final wave-function
$\Psi({\bf r},T)$. We calculate the probability of
having the electron in the $nlm$ bound state by projecting
$\Psi({\bf r},T)$ onto the corresponding field-free eigenstates $\Psi_{nlm}({\bf r})$,
\begin{equation}
\label{e2} p_{nlm}=|<\Psi_{nlm}({\bf r})|\Psi({\bf r},T)>|^2.
\end{equation}
The probability of having the electron in the $n$-quantum state is $p_n=\sum_{lm}p_{nlm}$.

\subsection{MC}

Tunnel ionization dominates if the Keldysh parameter $\gamma=\sqrt{2I_p}\omega/F(t_0)<1$~\cite{Keldysh65}, where
$I_p$ is the ionization potential, $\omega$ is the carrier frequency of the laser field.
If the polarization direction of the laser electric fields is along the z-axis,
the trajectories start at time $t_0$ at the
tunnel exit with the coordinates
\begin{equation}
\label{epos}
\left\{  \begin{aligned}
z(t_0)&=\frac{I_p+\sqrt{I_p^2-4F(t_0)Z_c}}{2F(t_0)},\\
x(t_0)&=y(t_0)=0.
\end{aligned} \right.
\end{equation}
This expression requires $F(t_0)<I_p^2/4Z_c$, and the over-the-barrier ionization (OTBI) is avoided.
In accord with the tunneling model the initial momentum
in $z$ direction is zero, i.e., $p_z=0$. The probability $w_{\perp}$ of
tunneling with a certain lateral momentum
$p_{\perp}(t_0)=\sqrt{p_x^2(t_0)+p_y^2(t_0)}$ is given by~\cite{delone91}
\begin{equation}
\label{epperp}
w_{\perp}\propto |p_{\perp}|\exp{\Big(-p_{\perp}^2\frac{\kappa}{F(t_0)}\Big)},
\end{equation}
and the ionization probability $w_0$ is given by the ADK theory~\cite{Ammosov86},
\begin{equation}
\label{eadkrate}
w_0\propto \Big(\frac{2\kappa^3}{F(t_0)}\Big)^{2Z_c/\kappa-|m|-1}\exp{\Big(-\frac{2\kappa^3}{3F(t_0)}\Big)}.
\end{equation}
Here, $m$ is the magnetic quantum number, which is initially $m=0$, and $Z_c=1$ is the core charge.
$I_p=0.5$a.u. is the binding energy, and $\kappa=\sqrt{2I_p}$.

Using the probabilities in Eqs. (\ref{epperp}) and (\ref{eadkrate}) we randomly pick an initial
lateral momentum and an initial ionization time $t_0$.
The electron is then propagated by integrating Newton's equations, under the combined field of
laser field and Coulomb force,
\begin{equation}
\label{enewton}
\frac{d^2{\bf r}}{dt^2}=-{\bf F}(t)+\nabla(1/r).
\end{equation}
After the laser pulse, we evaluate the total energy $E=p^2/2-1/r$, where $p=\sqrt{p_z^2+p_{\perp}^2}$ is the momentum of the electron. If $E$ is negative, the electron is bounded and
we determine an effective principle quantum number $n_{eff}$ and an effective angular momentum number $l_{eff}$ from
\begin{equation}
\label{eeleff}
\left\{  \begin{aligned}
E &=-\frac{1}{2n_{eff}^2}, \\
|{\bf L}|^2 &=l_{eff}(l_{eff}+1),
\end{aligned} \right.
\end{equation}
where the classical angular momentum reads
${\bf L}= {\bf r}\times {\bf v}= (r_{\parallel}v_{\perp}-r_{\perp}v_{\parallel}){\bf e}_{\parallel}\times {\bf e}_{\perp}$.
To compare with the quantum results from TDSE calculations,
the probabilities with $n_{eff}$ and $l_{eff}$ are integrated within each unit interval.
If $v_{\parallel}=0$, we yield $l=n-1$, corresponding to a circular Rydberg state.

\section{CALCULATION}

\begin{figure}
\includegraphics*[width=3in]{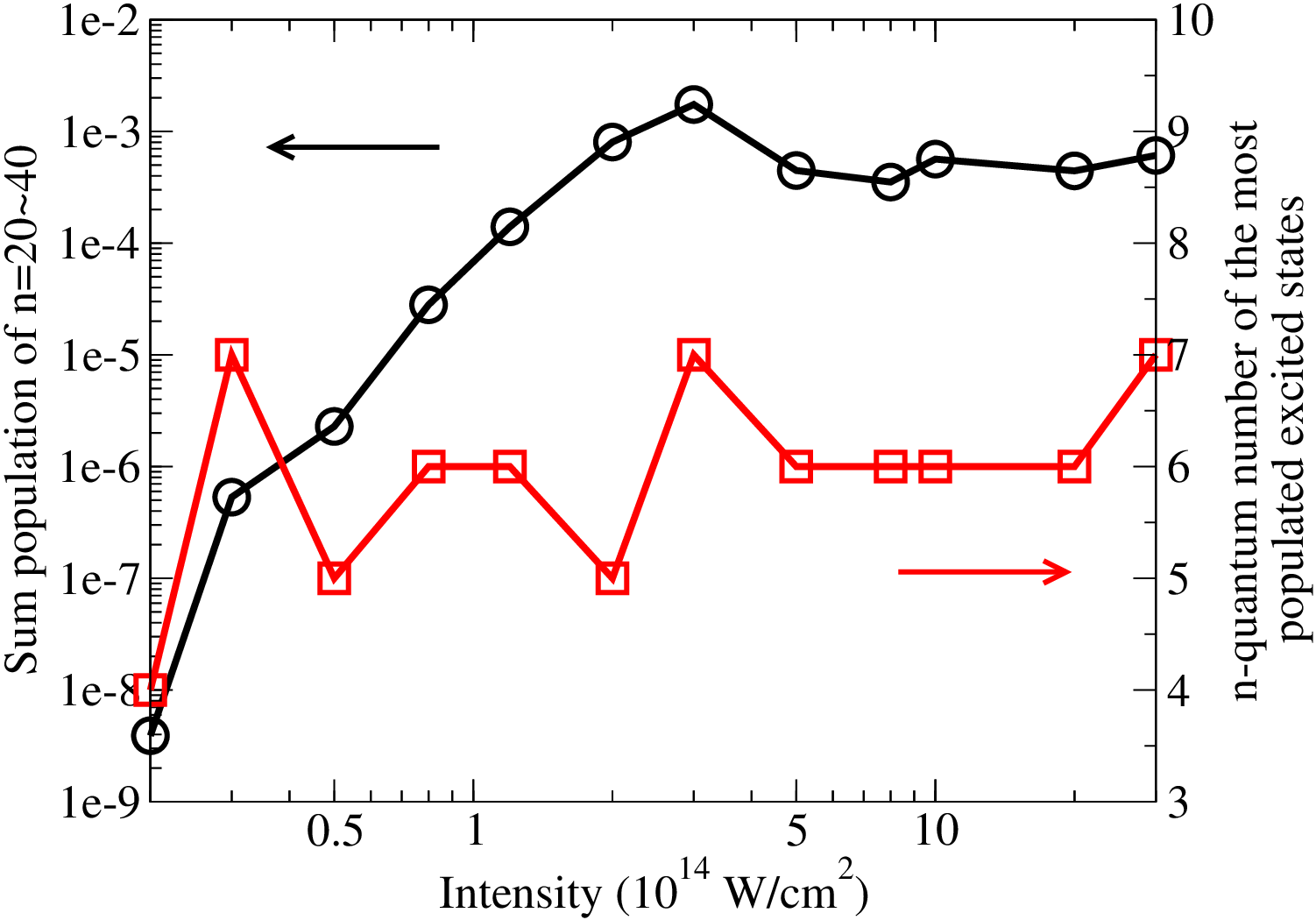}
\caption{(Color online)
Sum population of excited states of $n=20\sim 40$ ($\bigcirc$) and the $n$-quantum number of the most populated excited states ($\square$) versus the laser intensity, from
the TDSE calculations.
The laser pulse duration is 10 optical cycles, and the carrier wavelength is 800nm.
}\label{fmcpvsi}
\end{figure}

For the TDSE calculations, high $l$-quantum number should be used in the partial wave expansion in Eq.~(\ref{ewfexpand}),
if the Hydrogen atom is subjected to intense laser fields.
For the laser pulses we have used in this paper, convergency is reached at $l_{max}=80$.
For Hydrogen atoms, according to the classical Bohr-Sommerfeld model, the orbital radius scales as $n^2$ and $2n^2$ for $l=n-1$ and $l=0$ for the principle quantum number $n$~\cite{Gallagher88}.
In this paper, we investigate the Rydberg states generation of Hydrogen atoms with $n\le 40$, which
results in a simulation box as large as $r_{max}\approx 3000$ a.u.
The size of the Gauss-Radau quadrature $n_r=2000$.
For this large simulation box, the dipole approximation is still valid for the $800$nm laser fields.

The ground and excited states of Hydrogen atoms are calculated accurately by
dialogizing the field-free hamiltonian ($H_0$), and the energy of the ground state
$E_{1s}=-0.499999999998$a.u., with a relative error of $10^{-12}$ as compared to the exact value (0.5a.u.).
For the split-operator propagation scheme, the field-free propagator $\exp(-i\frac{1}{2}\Delta t H_0)$ needs only be
constructed once before the propagation, using the energy
values and eigenstates of the unperturbed system.
To improve the numerical stability of the propagation, a cutoff in the energy is applied to get rid of the spurious transitions
to the irrelevant regions of the very high energy spectrum.
The external field operator $\exp(-i\Delta t V(t))$ is diagonal in coordinate representation
when using the length gauge.
The time step for the propagation takes $\Delta t=0.01$a.u.
For efficient matrix and
vector operations we use the basic linear algebra subroutines
(BLAS)~\cite{blas} and the linear algebra package (LAPACK)~\cite{lapack}.

For the MC simulations, about $2\times 10^6$ trajectories have been launched every optical cycle at each fixed laser intensity, randomly varying the initial lateral momentum and ionization time.
To investigate the effects of Coulomb force and initial lateral momentum,  we have also performed MC simulations:
(a) ignoring the Coulomb force after the electron tunnels out the barrier [$V_c(t)=0,t>t_0$];
(b) ignoring the initial lateral momentum distribution at the tunnel exit [$p_{\perp}(t_0)=0$].
Comparisons and discussions with TDSE calculations are given.

The multi-cycle laser electric field ${\bf F}(t)$ is chosen as ${\bf F} (t)=F_0{\bf e}_F\sin^2(\pi t/\tau)\sin(\omega t+\delta)$, where $F_0$ is the peak field amplitude, $\tau$ is the pulse duration and $\delta$ is the carrier envelope phase (CEP).
In this paper, the unit of laser intensity $I_0=1\times 10^{14}$W/cm$^2$.
We use multi-cycle pulses with a pulse duration of $10T$, where $T=2\pi/\omega$ is the optical period.
For a HCP pulse, the electric field is ${\bf F} (t)=F_0{\bf e}_F \sin(\omega t+\delta)$, with $0\le t\le \tau=T/2$.
The phase $\delta$ is used to control the parity of the unipolar field.

\begin{figure}
\includegraphics*[width=3in]{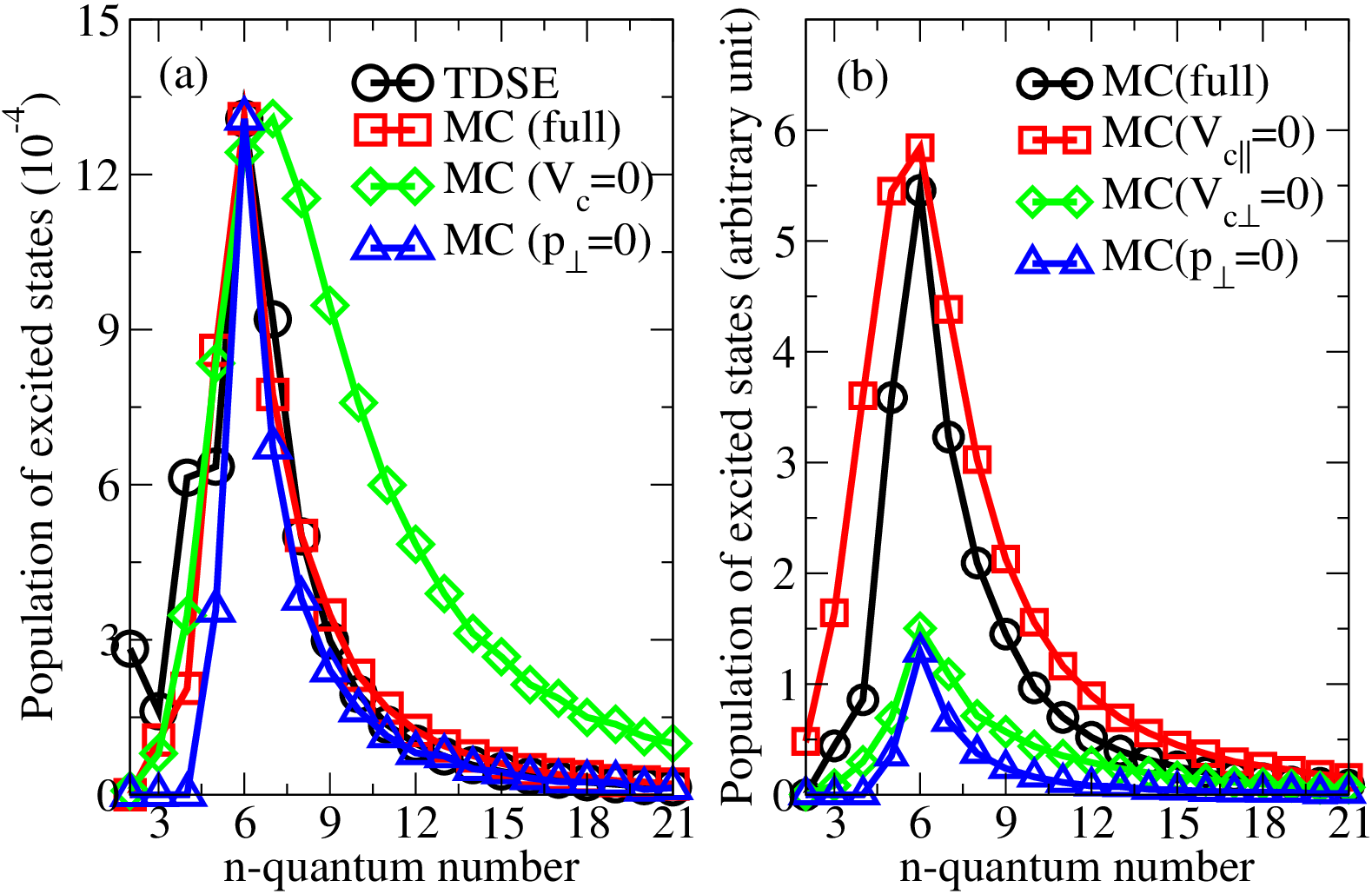}
\caption{(Color online)
(a) $n$-distribution of the population of excited states from
a TDSE calculation ($\bigcirc$) and MC simulations: full($\square$); $V_c=0$($\Diamond$); $p_{\perp}=0$($\triangle$). The populations from MC simulations have been normalized to maximum of the TDSE calculation at $n=6$.
(b) $n$-distribution of the population of excited states from
MC simulations: full($\bigcirc$); $V_{c\parallel}=0$($\square$); $V_{c\perp}=0$($\Diamond$); $p_{\perp}=0$($\triangle$).
The laser pulse duration is 10 optical cycles, and the intensity is $1.2\times 10^{14}$W/cm$^2$.
}\label{fmcpdis}
\end{figure}

\section{RESULTS AND DISCUSSIONS}

Fig.~\ref{fmcpvsi} presents the sum population of excited states of $n=20\sim 40$ and the $n$-quantum number of the most populated excited states for different laser intensities, from TDSE calculations.
The laser pulse duration is 10 optical cycles.
The sum population increases rapidly for relatively weak laser intensities, and reaches a plateau at higher intensities.
Ionization saturation is reached at higher laser intensities, while the population of highly excited Rydberg states does not drop (in fact,
there is a slight increase at a intensity of $10^{15}$W/cm$^2$ and higher).
This plateau of population may be explained by the fact that highly excited Rydberg atoms can survive in very intense laser fields,
as the experimental investigation of Helium atoms~\cite{Eichmann13}.
For the non-resonant $800$nm laser fields we have used, the $n$-quantum number of the most populated excited states
is insensitive to the laser intensity, yielding a principle quantum number of 6 in average.

To investigate the $n$-distribution of the population of excited states in detail, Fig.~\ref{fmcpdis}(a) presents the results from
TDSE calculation and MC simulations, at a laser intensity of $1.2\times 10^{14}$W/cm$^2$.
This intensity is chosen so that tunnel ionization is satisfied while OTBI
is avoided.
For comparison, the MC results are normalized according to the TDSE result.
For the TDSE calculation, a sharp maximum in the distribution is found around $n=6$.
The full MC simulation yields a result in good agreement with TDSE.
If the nuclear Coulomb force is ignored in the classical propagation ($V_c=0$), MC simulation yields
a broad distribution in the high $n$-quantum number part, in qualitative disagreement with TDSE
and full MC results.
This demonstrates the importance of the Coulomb force in the generation of highly excited
Rydberg states.
On the other hand, if the initial lateral momentum distribution is set to zero ($p_{\perp}=0$), the result still seems to be in qualitative agreement with TDSE and full MC simulations.

\begin{figure}
\includegraphics*[width=3in]{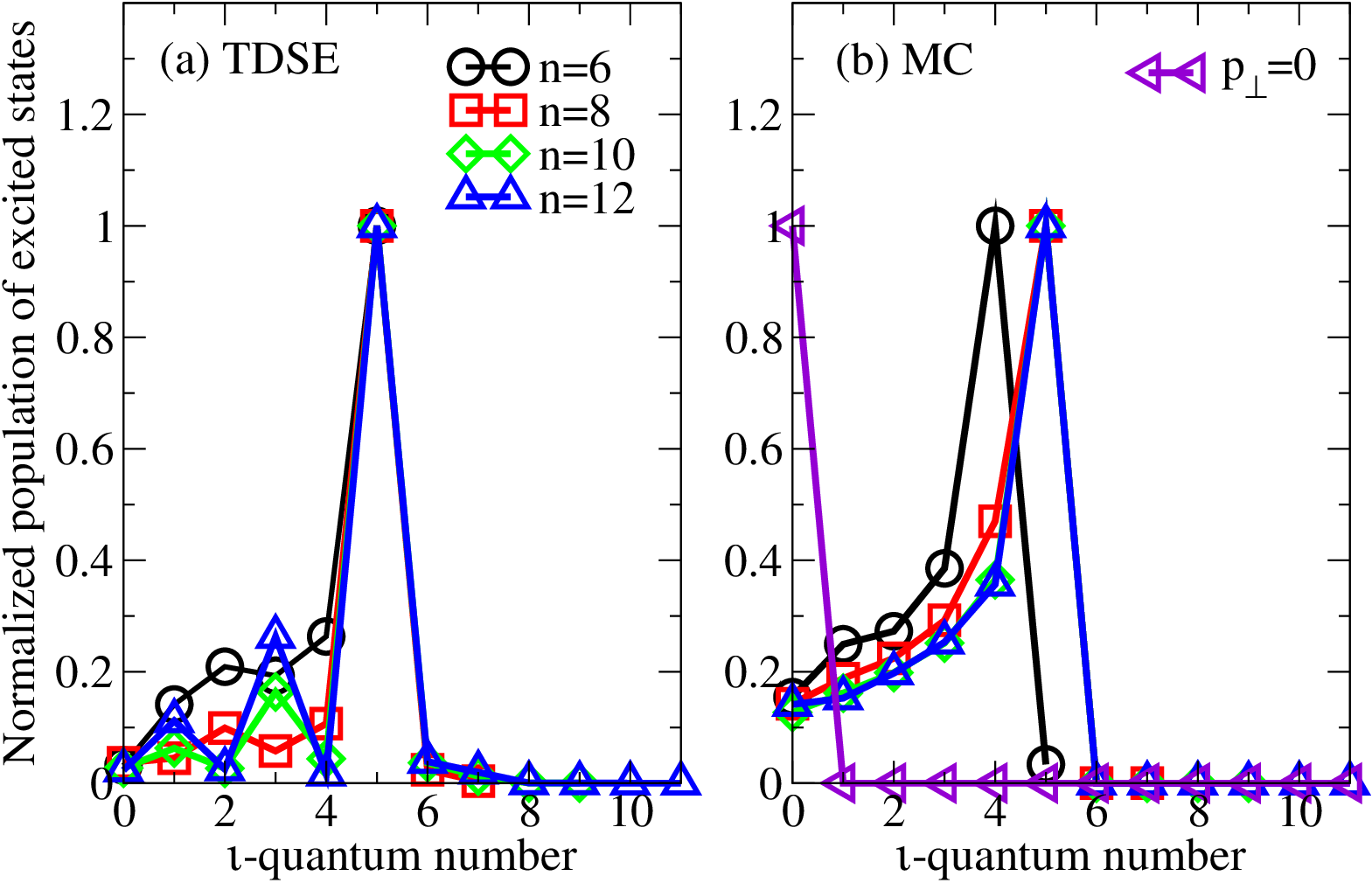}
\caption{(Color online)
Normalized $l$-distribution of the population of excited states, when the $n$-quantum number take: $n=6$($\bigcirc$), $n=8$($\square$), $n=10$($\Diamond$) and $n=12$($\triangle$),
from TDSE calculations (a) and full MC simulations (b), respectively.
The typical $l$-distribution from MC simulations with $p_{\perp}=0$($\lhd$)
is also presented in (b) for comparison.
The laser parameters are the same as in Fig.~\ref{fmcpdis}.
}\label{fmcpdis2}
\end{figure}

Does the initial lateral momentum have little effects in this case?
To answer this, in contrast to the normalized results in Fig.~\ref{fmcpdis}(a), Fig.~\ref{fmcpdis}(b) presents the unormalized population results from different MC simulations.
For the linearly polarized laser fields, the Coulomb force can be divided into two parts: the forces parallel ($V_{c||}$) and perpendicular ($V_{c\perp}$) to the polarization direction of the laser electric field.
If the parallel component $V_{c||}$ is set to 0, the change in the population is small as compared to the full result.
However, if the perpendicular component $V_{c\perp}$ is set to 0, the population is greatly suppressed.
This originates from the Coulomb focusing effect~\cite{brabec96,comtois05}.
During the propagation, the amplitude of lateral momentum decreases due to Coulomb focusing,
which focuses parts of the electron wave function, increasing the efficiency of rescattering.
This population suppression shows that the initial lateral momentum is important since the perpendicular force acts only on the lateral momentum.
On the other hand, by setting the initial lateral momentum $p_{\perp}$ at 0, a similar suppression to the case of $V_{c\perp}=0$ is observed in the population, which demonstrates that in the full MC simulation the dominant contribution comes from electrons with near-but-none zero initial lateral momentum.
Upon rescattering, the none-zero lateral momentum will results in a none-vanishing
angular momentum [see Eq.~(\ref{eeleff})].
To check this, the $l$-dependent populations of excited states
are compared in Fig.~\ref{fmcpdis2}.
In experiment, the $l$-distributions can be measured with the $l$-state selective field ionization~\cite{Gurtler04}.
The $n$-quantum number takes 6, 8, 10, and 12, respectively.
The MC simulation with zero $p_{\perp}$ yields only the $l=0$ states, and in (b) a single line is presented
for $p_{\perp}=0$.
For comparison, the results are normalized to the maximum.
As our expectation, for the TDSE and full MC simulations, the states with none-zero $l$ numbers exist and even overrun the $l=0$ part.
Note that due to the Coulomb focusing, the distributions in the $l$-quantum number center mostly at the smaller $l$ part.

\begin{figure}
\includegraphics*[width=3in]{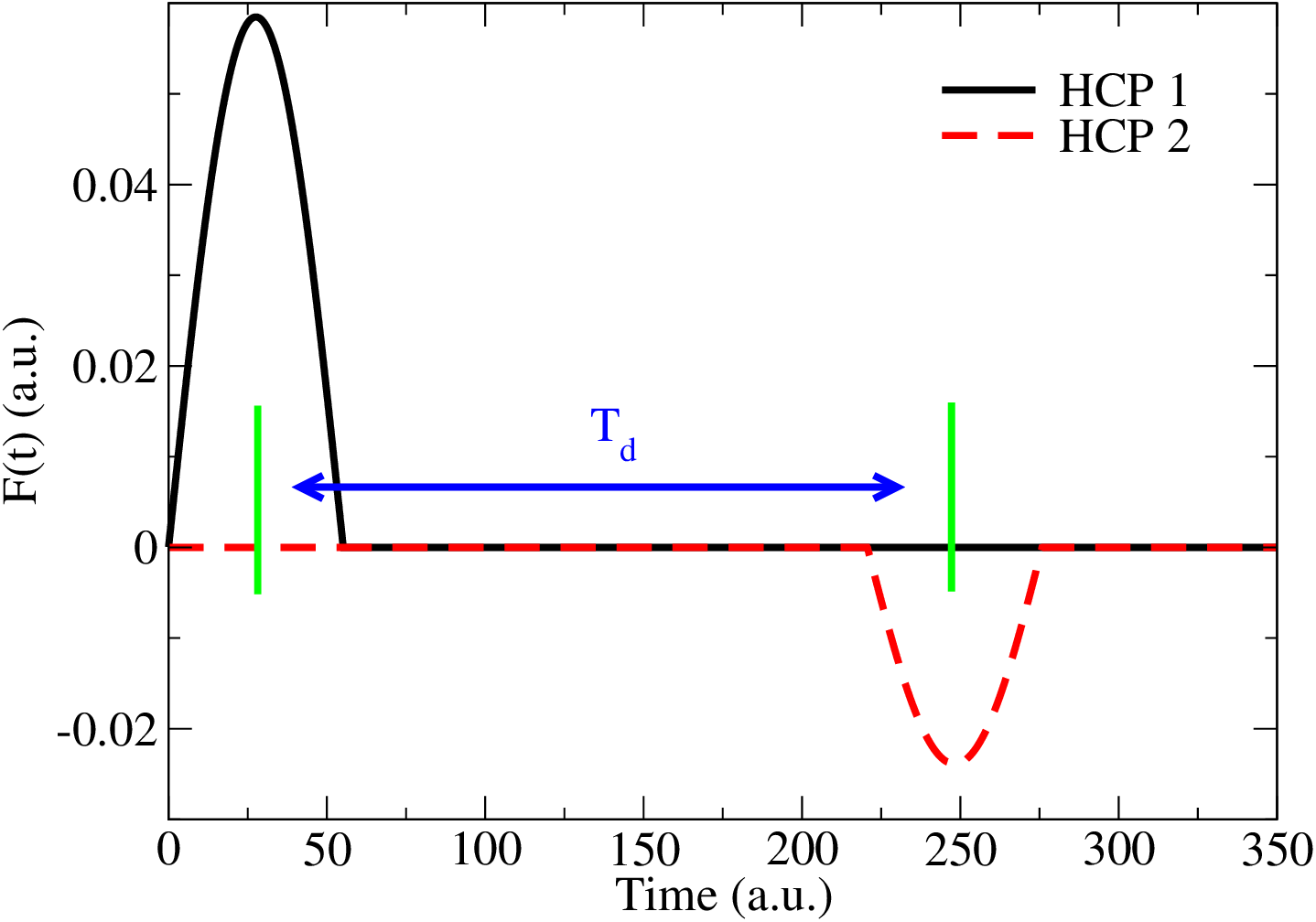}
\caption{(Color online)
Visualization of two HCP pulses with a time delay $T_d$ between the peak electric fields.
The intensities of the two pulses are $1.2\times 10^{14}$W/cm$^2$ and $0.2\times 10^{14}$W/cm$^2$, respectively.
The time delay in this figure is $T_d=2$T.
}\label{fhcpel}
\end{figure}

The above investigations show that both the Coulomb force and initial later momentum are important
in the generation of excited states.
The initial lateral momentum affects the $l$-distribution of population,
and the Coulomb force plays an important role in the lateral direction.
But how about the Coulomb force in the parallel direction?
These effects may be stronger for higher lying Rydberg states, while the $n$-quantum number of the most populated Rydberg states centers at small values (typically less than 10) for multi-cycle laser pulses.
For Helium atoms, the $n$-distribution of the population of excited states
has a similar sharp peak with a maximum around $n=8$, at an intensity of $10^{15}$W/cm$^2$~\cite{Nubbemeyer08}.
Also, the dynamics of electron is complicate for multi-cycle pulses, making it difficult to
investigate the physical mechanism for the formation of Rydberg states.
In the following the more simplified tool of HCP fields are utilized.

\begin{figure}
\includegraphics*[width=3in]{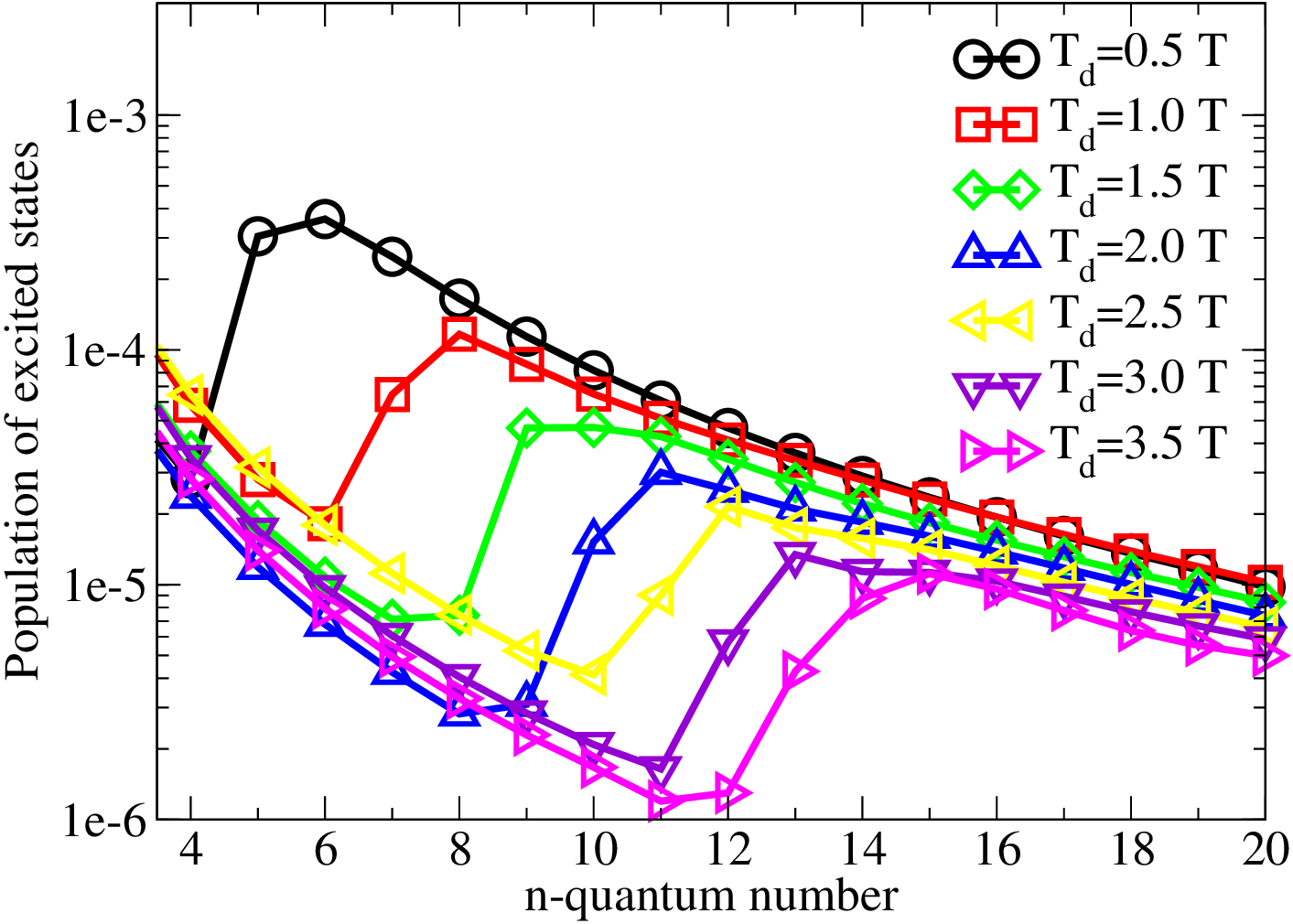}
\caption{(Color online)
$n$-distribution of the population of excited states from TDSE calculations with two time-delayed HCPs.
The intensities of the two pulses are $1.2\times 10^{14}$W/cm$^2$ and $0.2\times 10^{14}$W/cm$^2$, respectively.
The time delay $T_d$ is in unit of optical cycles, varying from 0.5 to 3.5.
}\label{fhcpdis}
\end{figure}

To generate very high lying Rydberg states, we have utilized a pair of HCPs.
The HCP pair is visualized in Fig.~\ref{fhcpel}.
A positive field HCP (HCP 1) is followed by a negative field HCP (HCP 2),
with a time delay $T_d$ between the peak electric fields.
HCP 1 serves as a "pump" field, which kicks the electron in the ground state and initiates an outgoing electron wavepacket.
The "probe" field HCP 2 kicks the outgoing electron in the reverse direction and traps it in the
high lying excited states.
Fixing the maximum intensities of the two pulses at $1.2\times 10^{14}$W/cm$^2$ and $0.2\times 10^{14}$W/cm$^2$ while varying the time delay, the $n$-distributions of the population of excited states from TDSE calculations are presented in Fig.~\ref{fhcpdis}.
As $T_d$ increases, the peak in the distribution moves to higher $n$-quantum
states.
For $T_d=0.5$T, the head and rear of these two HCPs meet and a complete optical
cycle is formed.
The peak is found at $n=6$, in agreement with the long-pulse case (see Fig.~\ref{fmcpdis}).
The position of the peak shifts to $n=15$ at $T_d=3.5$T.
To verify this kicking-trapping scenario, we have also perform simulations employing only the first HCP.
The peaks in the distribution disappear in this single HCP case, showing that the electron is really trapped by HCP 2.
Using both the two HCPs, but projecting out all the bound states ($E<0$) after the interaction of HCP 1, the distribution is nearly the same as the full simulation.
The continuum wave function initiated by HCP 1 is responsible for the generation of excited states.

\begin{figure}
\includegraphics*[width=3in]{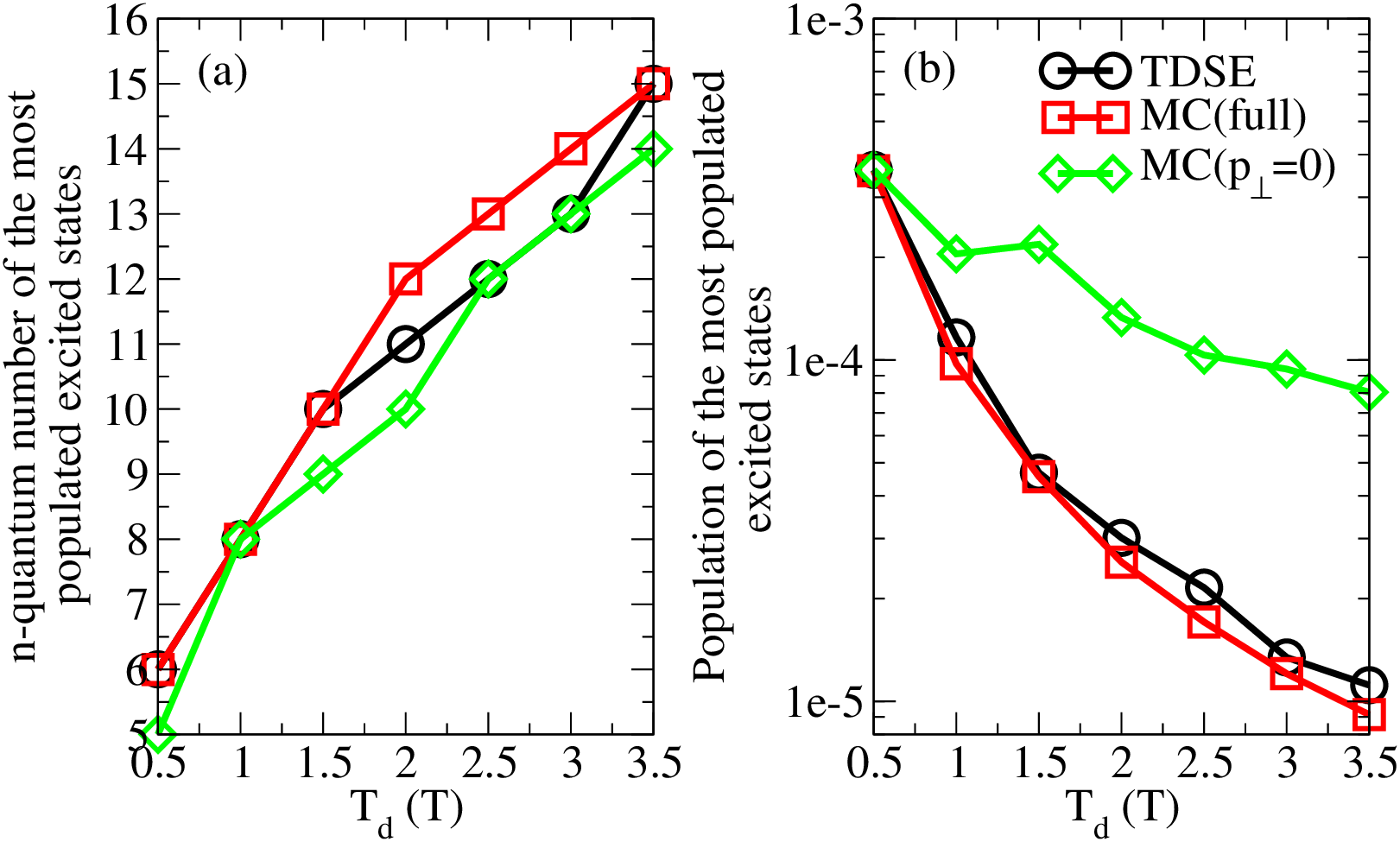}
\caption{(Color online)
The $n$-quantum number of the most populated Rydberg states (a) and the corresponding populations (b), versus the time delay $T_d$
between two HCPs,
from TDSE calculations ($\bigcirc$) and MC simulations with($\square$) and without ($\Diamond$) the lateral momentum distribution.
The populations from MC simulations have been normalized to the TDSE calculation at $T_d=0.5$T in (b).
The laser parameters are the same as in Fig.~\ref{fhcpdis}.
}\label{fnmax}
\end{figure}

The $n$-quantum number of the most populated Rydberg states and the corresponding populations using these HCPs from MC simulations are compared in Fig.~\ref{fnmax}.
The TDSE results, which are extracted from Fig.~\ref{fhcpdis}, are also given for comparison.
Both the $n$-quantum number and population from full MC simulations are in good
agreement with TDSE results.
The shift in the maximum position can be explained with the energy change due to the interaction of the ionized electron with the second HCP.
The energy change upon HCP 2 reads~\cite{Wetzels02}
\begin{equation}
\label{ehcpe}
\Delta E={\bf p}_t\cdot \Delta {\bf p}+\Delta p^2/2.
\end{equation}
where the momentum kick $\Delta {\bf p}=-\int {\bf F}(t)dt$.
For a fixed kick momentum $\Delta {\bf p}$, $\Delta E$ dependents on the electron momentum ${\bf p}_t$ at the instinct of kicking time $t$, which however depends on the distance $r_t$ between the outgoing electron and the nucleus.
During the field-free propagation between the two HCP fields (see Fig.~\ref{fhcpel}), due to the energy conservation $E_0=p_t^2/2-1/r_t$, the farther the distance $r_t$ is, the smaller the momentum $|{\bf p}_t|$ will be, leading to the shift in the most populated Rydberg states.
The Coulomb force is weak in the asymptotic region ($r\gg 0$), which is ignored in the classical propagation of the three-step model~\cite{Corkum93}.
However, our results show that the generation of highly excited Rydberg states is sensitive to the Coulomb field, even in the asymptotic region, due to the small energy difference between adjacent Rydberg states ($\Delta E_{n}\propto 1/n^3\ll 0$, for $n\gg 1$).
In Fig.~\ref{fnmax}(b), the population at the peak distribution decreases with increasing the time delay $T_d$.
This originates from the lateral wavepacket dispersion.
We turn to the MC simulations with $p_{\perp}=0$.
In Fig.~\ref{fnmax}(a), the MC simulations with $p_{\perp}=0$ yield qualitative agreement results with TDSE and full MC calculations.
The agreement is due to fully incorporation of the Coulomb force in the parallel direction.
In Fig.~\ref{fnmax}(b), for the MC simulation with $p_{\perp}=0$, although the maximum population still displays a slight decay, the decaying rate is much smaller than the TDSE and full MC cases, where lateral dispersion is considered.

\begin{figure}
\includegraphics*[width=3in]{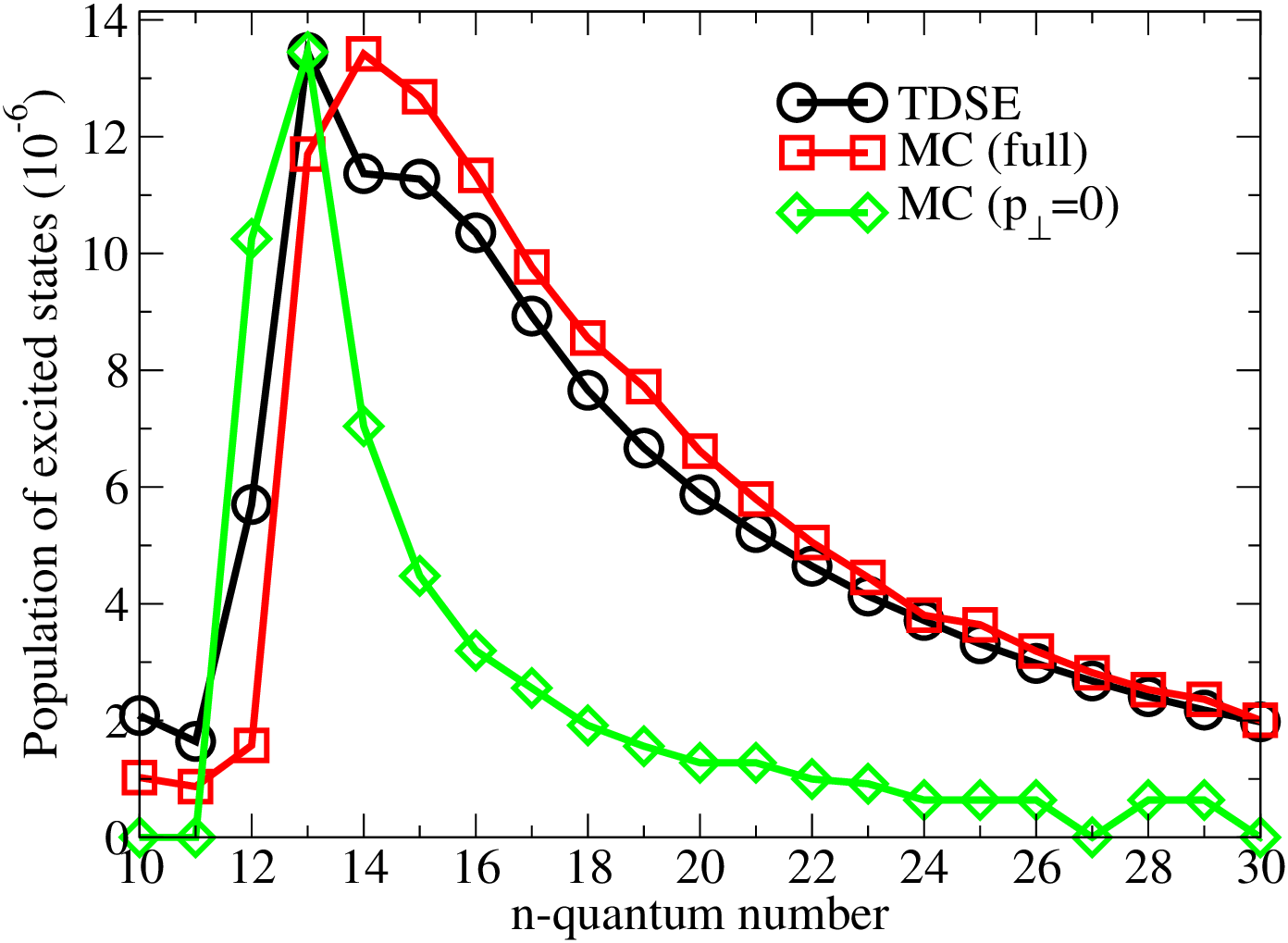}
\caption{(Color online)
$n$-distribution of the population of excited states from
a TDSE calculation ($\bigcirc$), MC simulations with ($\square$) and without ($\Diamond$) the initial transverse momentum distribution.
The populations from MC simulations have been normalized to maximum of the TDSE calculation at $n=13$.
The intensities of the two pulses are $1.2\times 10^{14}$W/cm$^2$ and $0.2\times 10^{14}$W/cm$^2$, respectively, with a time delay of $T_d=3.0$T.
}\label{fndis3}
\end{figure}

The lateral momentum plays an important role, not only
in the absolute population of the most populated Rydberg states, but also in the normalized $n$-distribution of the population as well.
For example, in Fig.~\ref{fndis3} we present the $n$-distribution of the population of excited states at a time delay of $T_d=3.0$T.
The TDSE result is extracted from Fig.~\ref{fhcpdis}.
The MC simulation with $p_{\perp}$ predicts a narrow
distribution, similar to the MCP case except for a shift in the position of the peak (Fig.\ref{fmcpdis}).
However, the TDSE and full MC calculations predict much broader distributions,
especially for the highly excited states.
This is different from the MCP case where the TDSE and full MC calculations yield narrow
distributions as well.
The agreement between TDSE and full MC results demonstrates the accuracy
of the initial lateral momentum distribution given by the ADK tunneling theory [Eq.~(\ref{epperp})].
The width of the $n$-distribution may be used as a tool to measure the
lateral momentum distribution for the tunnel ionization.

\begin{figure}
\includegraphics*[width=3in]{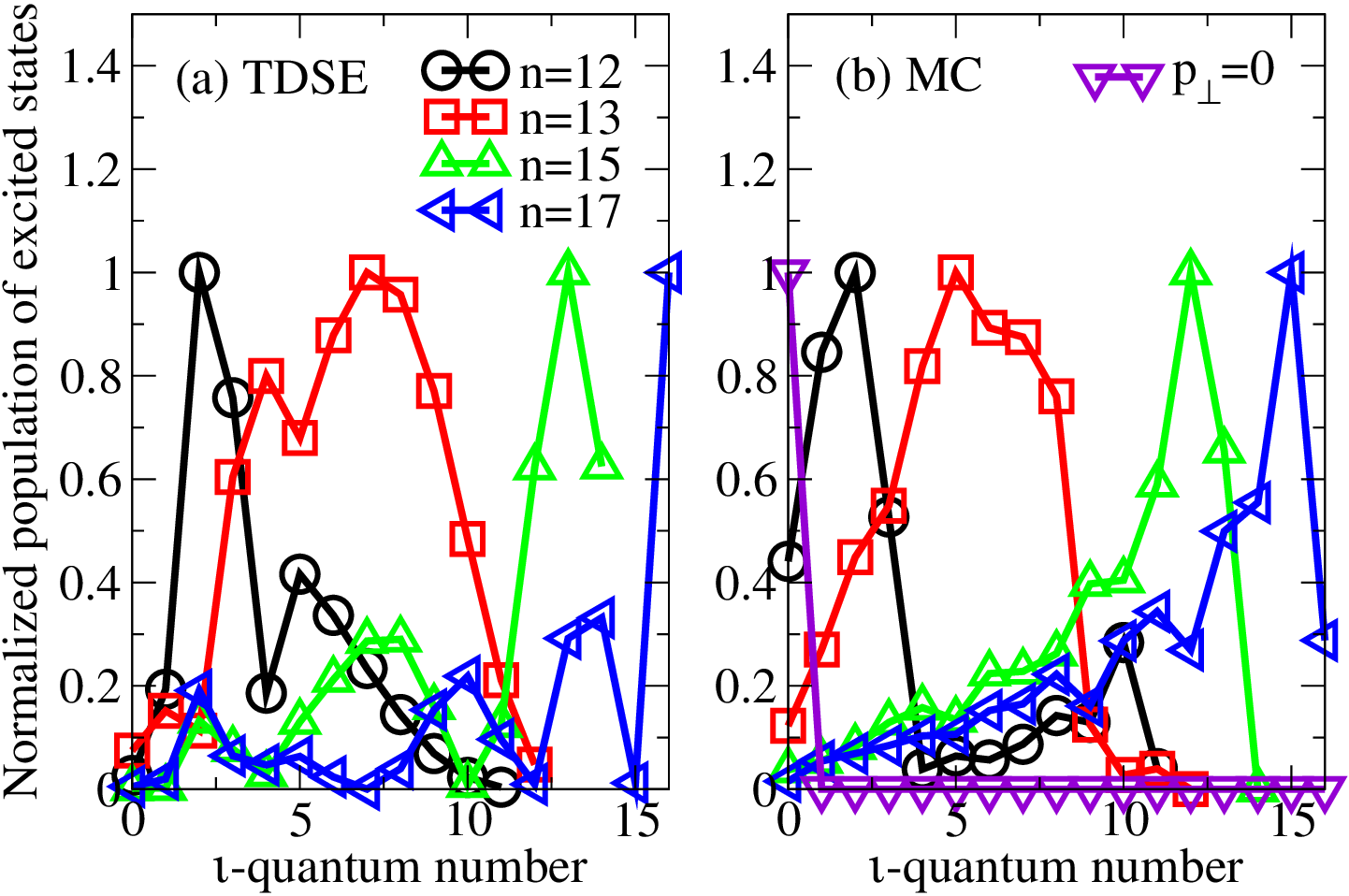}
\caption{(Color online)
Normalized $l$-distribution of the population of excited states,
when the $n$-quantum number take: $n=12$($\bigcirc$), $n=13$($\square$) and $n=15$($\triangle$),
from TDSE calculations (a) and full MC simulations (b), respectively.
The typical $l$-distribution from MC simulations with $p_{\perp}=0$($\lhd$)
is also presented in (b) for comparison.
The laser parameters are the same as in Fig.~\ref{fndis3}.
}\label{fldis3}
\end{figure}

To investigate how the lateral momentum broadens the $n$-distribution in detail, we calculate the $l$-distributions for several $n$-quantum numbers, as presented in Fig.~\ref{fldis3}.
A gradual change in the $l$-distributions is observed with increasing $n$.
For $n=12$, which lies on the rapid rising edge before the peak distribution,
the major contribution comes from the low $l$-quantum states ($l_{max}=2$).
For the peak distribution at $n=13$, the intermediate $l$-quantum states
contribute the most.
On the slowly decaying side, the high $l$ states dominates the distribution for $n=15$ (and for $n>15$).
The full MC simulations predict qualitative agreement results with TDSE.
Setting the initial lateral momentum to zero, only the $l=0$ state
contribute for each $n$-quantum number.
This explains the difference of the distribution width in Fig.~\ref{fndis3}.
In contrast to the MCP case where the $l$-distribution is limited to the small $l$
states, the $l$-distribution can reaches high $l$ states in the HCP case.
In the MCP case, the Coulomb focusing effect is enhanced in the multiple returning case with multi-cycle
laser pulses~\cite{bhardwaj01}.
In the HCP case, however, the tunneled electron need not return to the vicinity of the nucleus.
Thus Coulomb focusing is weakened and high angular momentum states can be formed.
The Rydberg states with high $l$ values ($l\sim n$) can have
very long lifetimes ($\tau_l\propto n^3l^2$)~\cite{Gallagher88}.
With properly chosen parameters,
this HCP pair scheme may provide a universal way to selectively excite atoms into arbitrary $nl$ state.
The time delay between the two HCP fields controls the energy, so as the principle $n$-quantum number of the Rydberg states.
The lateral momentum is related to the impact parameter $b$, thus angular momentum ${\bf L}= {\bf b}\times {\bf v}$ and the $l$-quantum number can be determined.

\section{CONCLUSIONS AND OUTLOOK}
In conclusion, we have investigated the generation of Rydberg states of Hydrogen atoms with intense laser pulses.
The theoretical methods we use include the TDSE and classical-trajectory MC simulation.
For the multi-cycle pulses, the sum population of highly-excited Rydberg states increases rapidly at lower laser intensities, and
reaches a plateau at higher intensities due to the stabilization of Rydberg atoms in super intense laser fields.
The $n$-quantum number of the maximum population of excited states
is insensitive to the laser intensity.
A sharp maximum in the n-distribution of excited states from TDSE calculations is reproduced by the full MC simulations.
The initial lateral momentum is responsible for the non-zero $l$-quantum states,
and the Coulomb force plays an important role in the lateral direction due to Coulomb focusing effect.

For the half-cycle pulses, as the time delay between the two HCPs increases, the peak in the distribution shifts to higher $n$-quantum
states.
This originates from the energy conservation due to the Coulomb force in the asymptotic region, where the energy difference between adjacent Rydberg states is comparative to the Coulomb force.
The lateral wavepacket dispersion results in the decreasing of the population at the peak distribution.
The major contribution comes from the low $l$-quantum states for low $n$-quantum states, while the high $l$-quantum part dominates
the highly excited states.
High $l$-states are generated because of the weakening of Coulomb focusing in the asymptotic region.

Our single electron results obtained in this paper also have implications for the generation of Rydberg states in multielectron
atoms~\cite{Shomsky09}.
In this paper, the discussions are limited to the tunnel ionization regime.
With more intense laser fields, over-the-barrier regime will be reached and different characteristics as compared to the TI case may be expected.
The study in the over-the-barrier regime, which requires a microcanonical distribution~\cite{botheron09} instead of the one given by Eqs.~(\ref{epos}) and (\ref{epperp}), is in progress.

\begin{acknowledgments}
This work is supported by  the National Basic Research Program of China (973 Program)
under Grant No. 2013CB922203, the NSF of China  (Grant No. 11374366) and the Major Research plan of NSF of China (Grant No. 91121017).
B. Z. is supported by  the Innovation Foundation of NUDT under Grant No. B110204
and the Hunan Provincial Innovation Foundation For Postgraduate
under Grant No. CX2011B010.
\end{acknowledgments}

%\begin{references}

%\end{references}

\end{document}